\documentclass[aps,twocolumn,groupedaddress,floatfix]{revtex4-1}
\usepackage{amssymb,amsmath}
\usepackage{graphicx}
\usepackage{subfigure}
\usepackage[english]{babel}
\usepackage{float}
\usepackage{color}

\usepackage{lipsum}
\usepackage{suffix}
\usepackage{mathtools}

\DeclarePairedDelimiterX\MeijerM[3]{\lparen}{\rparen}%
{\begin{smallmatrix}#1 \\ #2\end{smallmatrix}\delimsize\vert\,#3}

\newcommand\MeijerG[8][]{%
  G^{\,#2,#3}_{#4,#5}\MeijerM[#1]{#6}{#7}{#8}}

\WithSuffix\newcommand\MeijerG*[7]{%
  G^{\,#1,#2}_{#3,#4}\MeijerM*{#5}{#6}{#7}}
\begin{document}
\newcommand{\be}{\begin{equation}}
\newcommand{\ee}{\end{equation}}
\newcommand{\rojo}[1]{\textcolor{red}{#1}}

\title{The Two-Dimensional Fractional Discrete Nonlinear Schr\"{o}dinger Equation}

\author{Mario I. Molina}
\affiliation{Departamento de F\'{\i}sica, Facultad de Ciencias, Universidad de Chile, Casilla 653, Santiago, Chile}

\date{\today }

\begin{abstract} 
We study a fractional version of the two-dimensional discrete nonlinear Schr\"{o}dinger (DNLS) equation, where the usual discrete Laplacian is replaced by its fractional form that depends on a fractional exponent $s$ that interpolates between the case of an identity operator ($s=0$) and that of the usual discrete 2D Laplacian ($s=1$). This replacement leads to a long-range coupling among sites that, at low values of $s$, decreases the bandwidth and leads to quasi-degenerate states. The mean square displacement of an initially-localized excitation is shown to be ballistic at all times with a `speed' that increases monotonically with the fractional exponent $s$. We also compute the nonlinear modes and their stability for both, bulk and surface modes. The modulational stability is seen to increase with an increase in the fractional exponent. The trapping of an initially localized excitation shows a selftrapping transition as a function of nonlinearity strength, whose 
threshold increases with the value of $s$. In the linear limit, there persists a linear trapping at small $s$ values. This behavior is connected with the decrease of the bandwidth and its associated increase in quasi-degeneracy.
\end{abstract}

\maketitle

{\em Introduction}. Let us consider the discrete nonlinear Schr\"{o}dinger (DNLS) equation in 
$d$-dimensions\cite{uno,dos,tres}:
\be
i {d C_{\bf n}\over{d \tau}} +  \sum_{\bf m} C_{\bf m} + \gamma |C_{\bf n}|^2 C_{\bf n} = 0\label{eq:1}
\ee
where the sum is over the nearest-neighbor sites. Parameters $V$ and $\gamma$ represent the coupling between nearest neighbor sites and the nonlinear coefficient, respectively. The DNLS equation has proven useful in describing  a variety of phenomena in nonlinear Physics, such as  propagation of excitations in a deformable medium\cite{davidov,christiansen}, dynamics of Bose-Einstein condensates inside coupled magneto- optical traps\cite{oberthaler,brazhniy}, transversal propagation of light in waveguide arrays\cite{demetrio,lederer,vicencio,jason}, self-focusing and collapse of Langmuir waves in plasma physics\cite{zakharov1,zakharov2} and description of rogue waves in the ocean\cite{onorato}, among others. Its main features include the existence of localized nonlinear solutions with families of stable and unstable modes, the existence of a selftrapping transition\cite{molina,tsironis} of an initially localized excitation, and a degree of excitation mobility in 1D\cite{vicencio2}. All these characteristics have made the DNLS into a paradigmatic equation that describes the propagation of excitations in a nonlinear medium under a variety of different physical scenarios.

Recently, the topic of fractional derivatives has gained increased attention. It started with the observation that a usual integer-order derivative could be extended to a fractional-order derivative, that is, $(d^n/dx^n)\rightarrow (d^s/dx^s)$, for real $s$, which is known as the fractional exponent. The topic has a long history, dating back to letters exchanged between Leibnitz and L'Hopital, and later contributions by Euler, Laplace, Riemann, Liouville, and Caputo to name some. In the Riemann-Liouville
formalism\cite{hermann,west,miller}, 
the s-th derivative of a function $f(x)$ can be formally expressed as
\be
\left({d^{s}\over{d x^{s}}}\right) f(x) = {1\over{\Gamma(1-s)}} {d\over{d x}} \int_{0}^{x} {f(x')\over{(x-x')^{s}}} dx'\label{eq:2}
\ee
for $0<s<1$. For the case of the laplacian operator $\Delta=\partial^2/\partial {\bf r}^2$, its fractional form $(-\Delta)^s$ can be expressed as\cite{landkof}
\be
(-\Delta)^s f({\bf x}) = 
L_{d,s}  \int { f({\bf x})-f({\bf y})\over{|{\bf x}-{\bf y}|^{d + 2 s}} }
\label{eq:3}
\ee
where,
\be
L_{d,s} = {4^s \Gamma[(d/2)+s]\over{\pi^{d/2} |\Gamma(-s)|}}
\ee
where $\Gamma(x)$ is the Gamma function, $d$ is the dimension,  and $0<s<1$ is the fractional exponent.
Fractional order differential equations constitute useful tools to articulate complex events and to model various physical phenomena. In particular, the fractional Laplacian (\ref{eq:3}) has found many applications in fields as diverse as 
fractional kinetics and anomalous diffusion\cite{71,86,101}, Levy processes in quantum mechanics\cite{75}, fluid mechanics\cite{30,35}, strange kinetics\cite{82}, fractional quantum mechanics\cite{64,65}, plasmas\cite{2}, biological invasions\cite{9} and electrical propagation in cardiac tissue\cite{20}.

In this work we examine the consequences of replacing the usual two-dimensional discrete Laplacian by its fractional form, focusing on the  effects on the existence and stability of nonlinear modes, as well as in the transport of excitations in a square lattice. In general, we find that the known DNLS phenomenology is more or less preserved, although there is a marked tendency towards band flattening as the fractional exponent $s$ decreases. This causes an increase in the system's degeneracy and affects its capacity to selftrap excitations.

{\em The model}. For a square lattice ($d=2$), the kinetic energy term $\sum_{\bf m}C_{\bf m}$ in Eq.(\ref{eq:1}), can be written as $4  C_{\bf n} + \Delta_{n} C_{\bf n}$ where $\Delta_{n}$ is the discretized Laplacian
\begin{eqnarray} 
\Delta_{n} C_{\bf n}&=& C_{p+1,q}+C_{p-1,q}-4\ C_{p,q}\nonumber\\
                    & &+ C_{p,q+1}+C_{p,q-1},
\end{eqnarray}
where ${\bf n}=(p,q)$. Equation (\ref{eq:1}) can then be written in dimensionless form as
\be
i {d C_{\bf n}\over{d t}} + 4 C_{\bf n} + \Delta_{n} C_{\bf n} + \chi |C_{\bf n}|^2 C_{\bf n} = 0.\label{eq:6}
\ee
\begin{figure}[t]
 \includegraphics[scale=0.2]{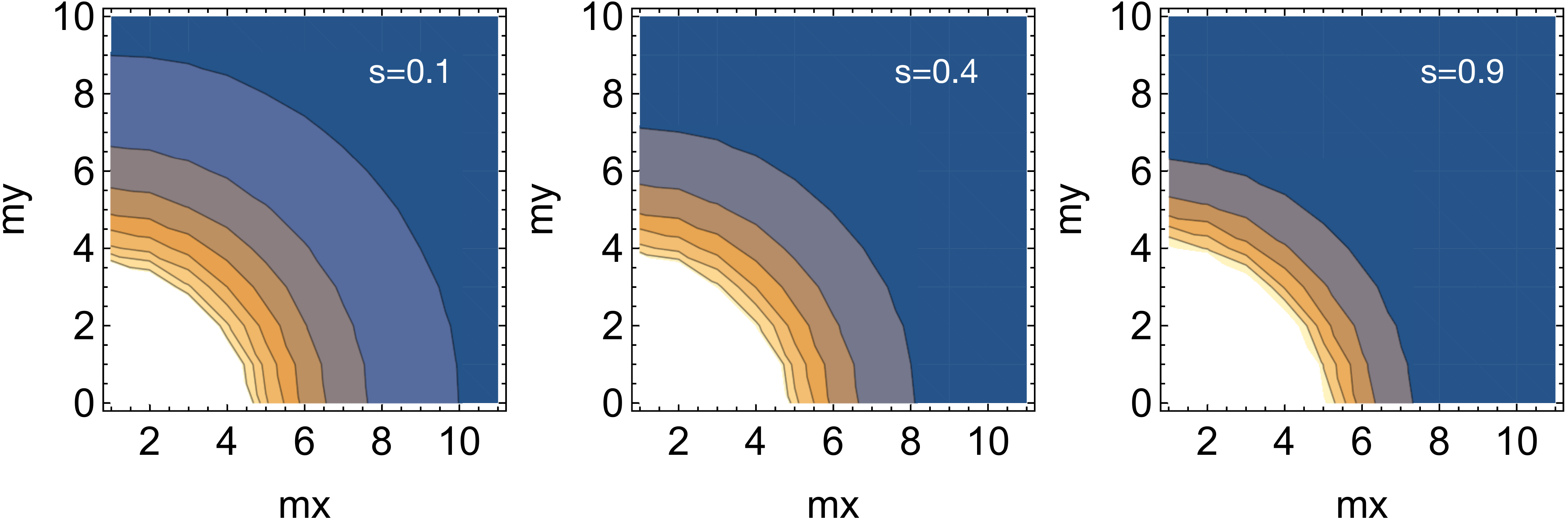}
  \caption{Effective coupling $K(m_{1},m_{2})$ for several fractional exponents. $N=11\times 11$}
  \label{fig1}
\end{figure}
where $t\equiv V \tau$ and $\chi\equiv \gamma/V$. We proceed now to replace $\Delta_{n}$ by its fractional form $(\Delta_{n})^s$ in Eq.(\ref{eq:6}). The form of this fractional discrete Laplacian for $d=2$ is given by\cite{discrete laplacian,luz}
\be 
(\Delta_{n})^s C_{\bf n} = \sum_{{\bf m}\neq{\bf n} } (C_{\bf m} - C_{\bf n}) \ K^{s}({\bf n}-{\bf m})\label{eq:7}
\ee
where,
\be 
K^{s}({\bf m}) = {1\over{|\Gamma(-s)|}}\ \int_{0}^{\infty} e^{-4 t }\ I_{m_1}(2 t)\ I_{m_2}(2 t)\ t^{-1-s}\ dt \label{eq:8}
\ee
with ${\bf m} = (m_1,m_2)$ and $I_{m}(x)$ is the modified special Bessel function. An alternative expression for $(\Delta_{n})^s$ is
\begin{widetext}
\be
(\Delta_{n})^s C_{\bf j} = 
                        L_{2,s}\sum_{{\bf m}\neq {\bf j}} (C_{\bf m}-C_{\bf j})\  
\MeijerG*{2}{2}{3}{3}{1/2,-(j2-m2+1+s,j2-m2+1+s)}{1/2+s,j1-m1,-(j1-m1)}{1}\label{eq:9}
\ee
\end{widetext}
where ${\bf j}=(j_1,j_2)$ and ${\bf m}=(m_1,m_2)$, and $G(...)$ is the  Meijer G-function. 
The symmetric kernel $K^{s}({\bf m})$ plays the role of a long-ranged coupling. Near $s=1$, $K({\bf m})\rightarrow \delta_{{\bf m},{\bf u}}$ where ${\bf u}=(1,0)$ or ${\bf u}=(0,1)$, i,e., coupling to nearest neighbors only. Let us look at its asymptotic form at long distances.  Using 
\be
I_{\nu}(z) \sim {1\over{2 \pi \nu}}\left( {e z\over{2 \nu}} \right)\hspace{1cm}\nu\rightarrow \infty.
\ee
plus $\Gamma(n)\sim n^{1/2}\ (n/e)^n$ and $\Gamma(n+s)\sim n^s\ \Gamma(n)$\ for $n\rightarrow \infty$, one obtains
\be
K^{s}({\bf m}) \sim \left( {m_{1}+m_{2}\over{m_{1} m_{2}}} \right) {4^{-(m_{1}+m_{2})}\over{|\Gamma(-s)|}} {(m_{1}+m_{2})^{m_{1}+m_{2}}\over{m_{1}^{m_{1}} m_{2}^{m_{2}}}}
\ee
where ${\bf m}=(m_{1}, m_{2})$. Thus, for the `diagonal' case $m_{1}=m_{2}=m$, one obtains
\be
 K^{s}(m) \sim {1\over{|\Gamma(-s)|}} {2^{-2 m}\over{\sqrt{m}}}\hspace{1cm} (m\rightarrow \infty).
\ee
This decay is faster than in the one-dimensional case\cite{molina1D}.
We consider now stationary modes defined by $C_{\bf n}(t)=e^{i \lambda t}\ \phi_{\bf n}$, which obey
\be
(-\lambda + 4 ) \phi_{\bf n} + \sum_{{\bf m}\neq {\bf n}} (\phi_{\bf m} - \phi_{\bf n}) K^{s} ({\bf m}-{\bf n}) + \chi |\phi_{\bf n}|^2 \phi_{\bf n} = 0 \label{stationary}
\ee
\begin{figure}[t]
 \includegraphics[scale=0.2]{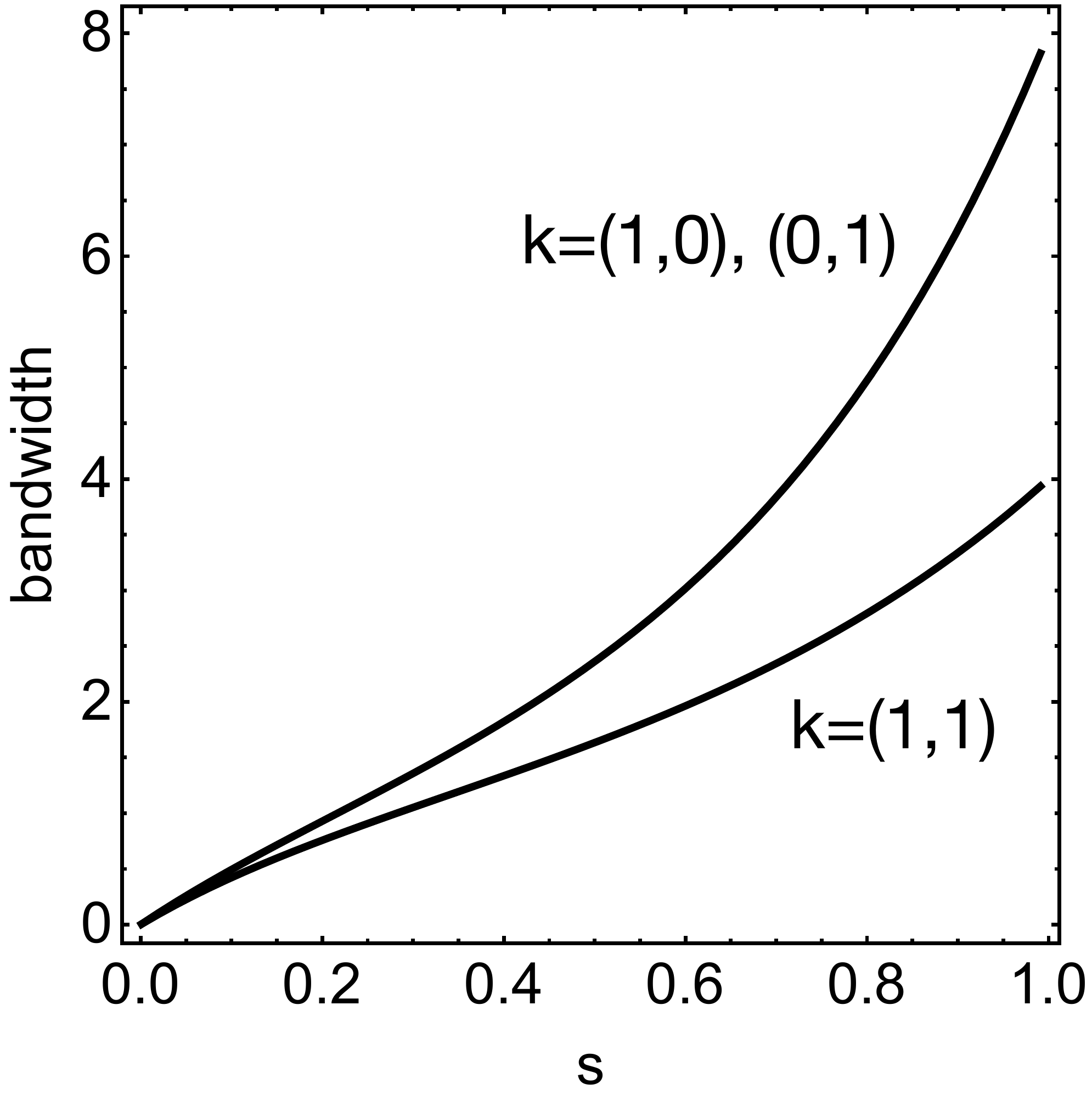}
 \includegraphics[scale=0.211]{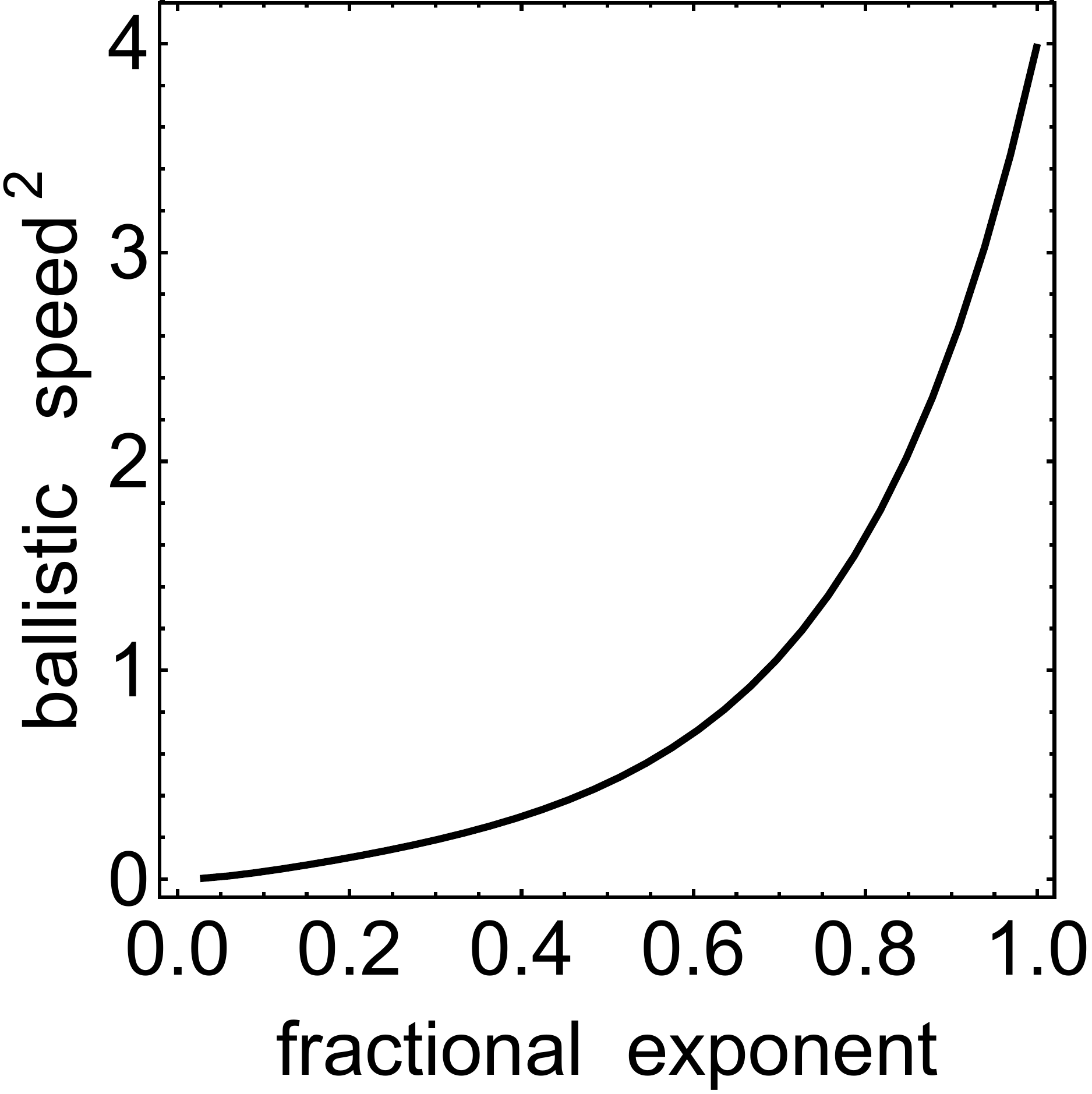}
 \caption{Left: Bandwidth along three $k$-space directions, as a function of the fractional exponent $s$. Right: Characteristic ballistic speed (square) as a function of the fractional exponent. Note that as $s\rightarrow 1$, this speed approaches $2$, as expected.
   ($N=11\times 11$).}
  \label{fig2}
\end{figure}
\begin{figure}[t]
 \includegraphics[scale=0.27]{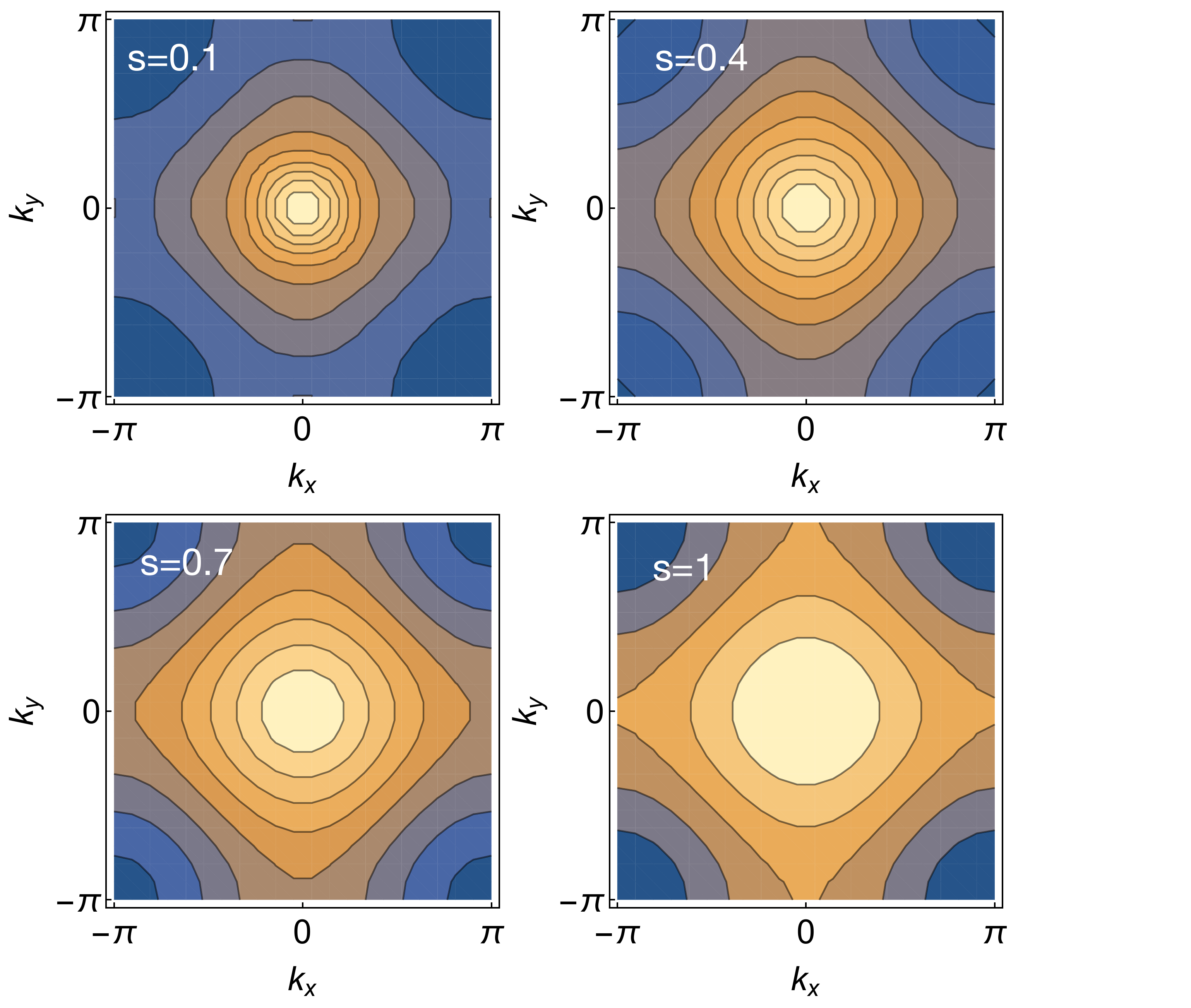}
  \caption{Contour plot of $\lambda(k_{x},k_{y})$ for several fractional exponents. $N=11\times 11$.}
  \label{fig3}
\end{figure}
\begin{figure}[t]
 \includegraphics[scale=0.2]{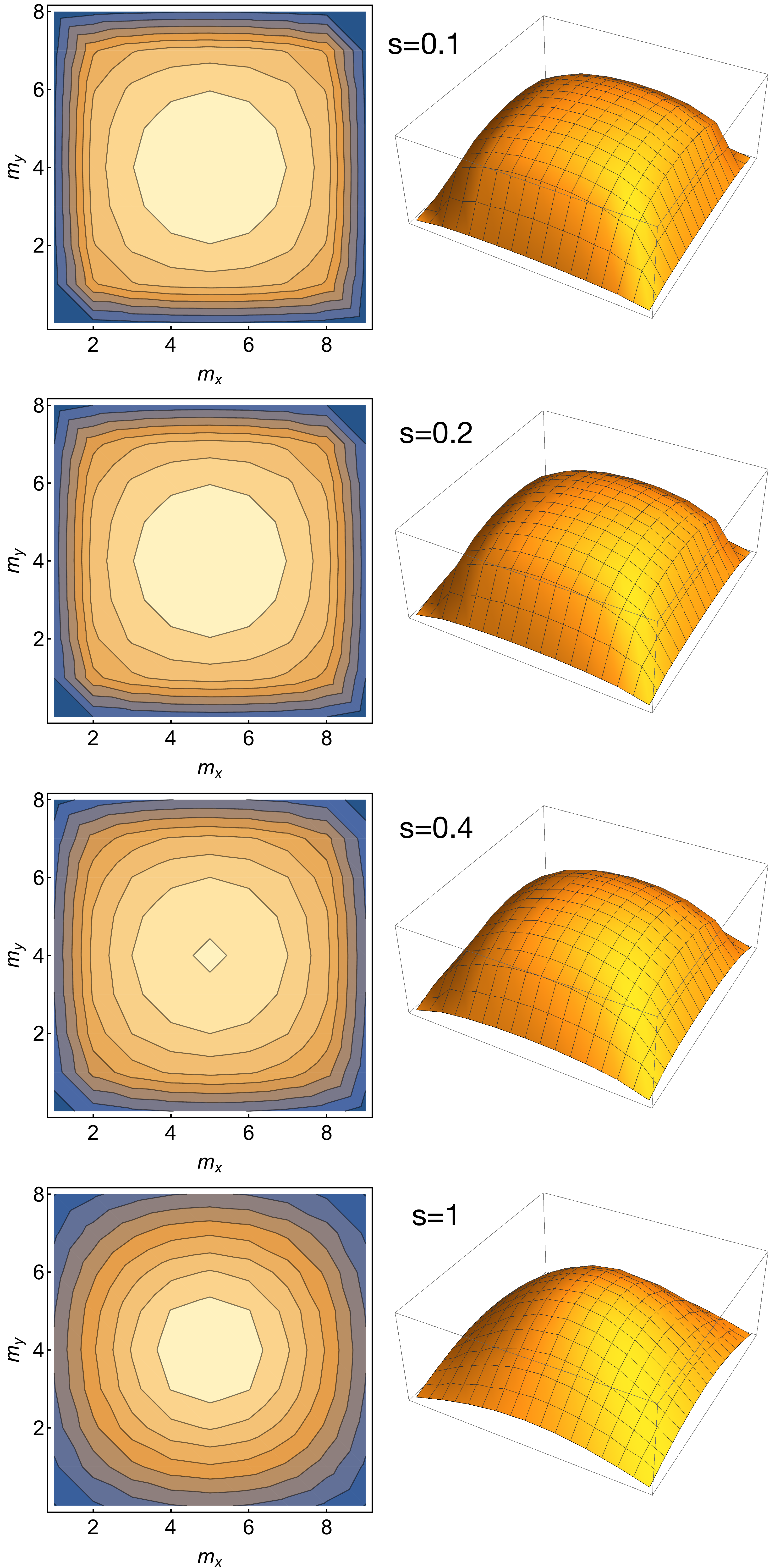}
  \caption{ Fundamental linear mode for several values of the fractional exponent. Left: Contour plots. Right: Three dimensional plots. $N=11\times 11$.}
  \label{fig4}
\end{figure}
It should be mentioned that in expressions (\ref{eq:6}) and (\ref{stationary}) the term $4$ is to be replaced by $3\ (2)$ for sites at the edge (corner), when dealing with a finite square lattice.
Figure \ref{fig1} shows $K({\bf m})$, where we see how the range of the coupling increases as $s$ decrease. This has the effect of increasing the coupling between distant sites, leading to deep consequences, as we will show below.

{\em Plane waves}. Let us set $\chi=0$ and look for plane wave solutions, $\phi_{\bf n} = A\ \exp(i {\bf k}.{\bf n})$, where we are assuming an infinite square lattice. After a short algebra, one obtains the dispersion relation,
\be
\lambda({\bf k}) = 4  + \sum_{\bf m} \left( e^{i {\bf k}.{\bf m}} -1 \right)\ K^{s}({\bf m})
\label{eq:11} 
\ee
Unfortunately, in this case it is not possible to rewrite Eq.(\ref{eq:11}) in closed form in terms of special functions, as in the one-dimensional case. Figure \ref{fig2} shows the bandwidth along two different directions in $k$-space. For both cases the bandwidth decreases monotonically as $s$ decreases. This flattening of the band increases the degeneracy of the modes, and the system becomes closer to an ideal model known as the `simplex'\cite{simplex1,simplex2} where every site is coupled to every other site with equal strength. This leads to a strong localization of an initially localized excitation. In our case, this will become evident when we look at selftrapping.

{\em Root mean square (RMS) displacement}. One of the ways to quantify the propagation of an excitation across a lattice, is by the Root Mean Square (RMS) displacement, defined as
\be
\langle {\bf n}^2 \rangle=\sum_{\bf n} {\bf n}^2\ |\phi_{\bf n}|^2/\sum_{\bf n}|\phi_{\bf n}|^2.\label{RMS}
\ee
A general result concerning RMS in lattices is that if $\lambda({-\bf k})=\lambda({\bf k})$, 
$\lambda({\bf k})=\lambda({\bf k}+2 \pi {\bf q})$ and $\nabla \lambda({\bf k})|_{FBZ}=0$,
where FBZ is the first Brillouin zone, then it can be proven that the RMS for an initially localized excitation ($\phi_{\bf n}(0)=\delta_{{\bf n},{\bf 0}}$) is always ballistic and given (in 2D) by\cite{martinez_molina}
\be
\langle {\bf n}^2 \rangle=\left[{1\over{4 \pi^2}} \int_{\mbox{\tiny{FBZ}}} \left( \nabla \lambda({\bf k}) \right)^2\ d\, {\bf k}  \right]\ t^2
\ee
where FBZ is the first Brillouin zone. Using the form of $\lambda({\bf k})$ given by Eq.(\ref{eq:11}), we obtain
\be
\langle {\bf n}^2 \rangle = \sum_{\bf m}\ {\bf m}^2\ K^{s}({\bf m}) K^{s}(-{\bf m})
\ee
The parity properties of the modified Bessel functions, imply that 
$K^{s}(-{\bf m})= K^{s}({\bf m})$. Therefore, 
\be
\langle {\bf n}^2 \rangle = \left[\  \sum_{\bf m} {\bf m}^2 ( K^{s}({\bf m}) )^2\ \right]\  t^2
\ee
where ${\bf m}=(m_{1}, m_{2})$. The quantity inside the square brackets can be interpreted as the square of a  characteristic ballistic speed. Given the rapid decrease of $K^{s}({\bf m})$ with distance, $\langle {{\bf n}^2} \rangle$ becomes well-defined for all $0<s<1$.
Figure 2b shows this speed square as a function of the fractional exponent $s$. The speed rises from zero at $s=0$ up to $4$ at $s=1$, which is the usual ballistic value. The vanishing of the speed at $s=0$ implies that, in this limit, the excitation is unable to move and remain localized at the initial site. In fact, at $s=0$ any initial condition, which is a combination of all plane waves, will be unable to diffuse way.
This result is in consonance with the observation that, at $s\rightarrow 0$, the band becomes completely degenerate, $\lambda({\bf k})\rightarrow 4$ causing the group velocity ${\bf \nabla} \lambda({\bf k})$ to vanish.

\begin{figure}[t]
 \includegraphics[scale=0.25]{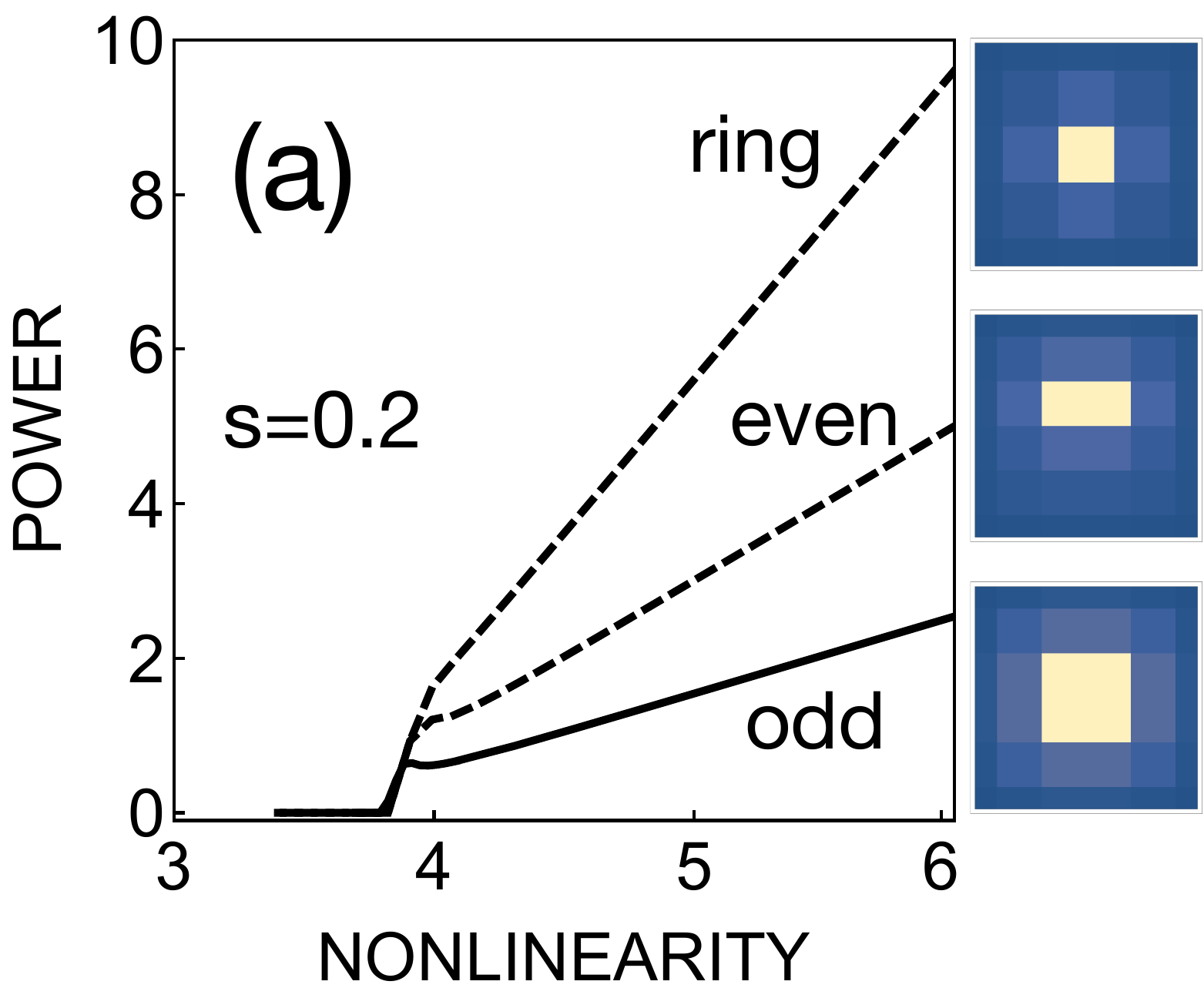}
 \includegraphics[scale=0.25]{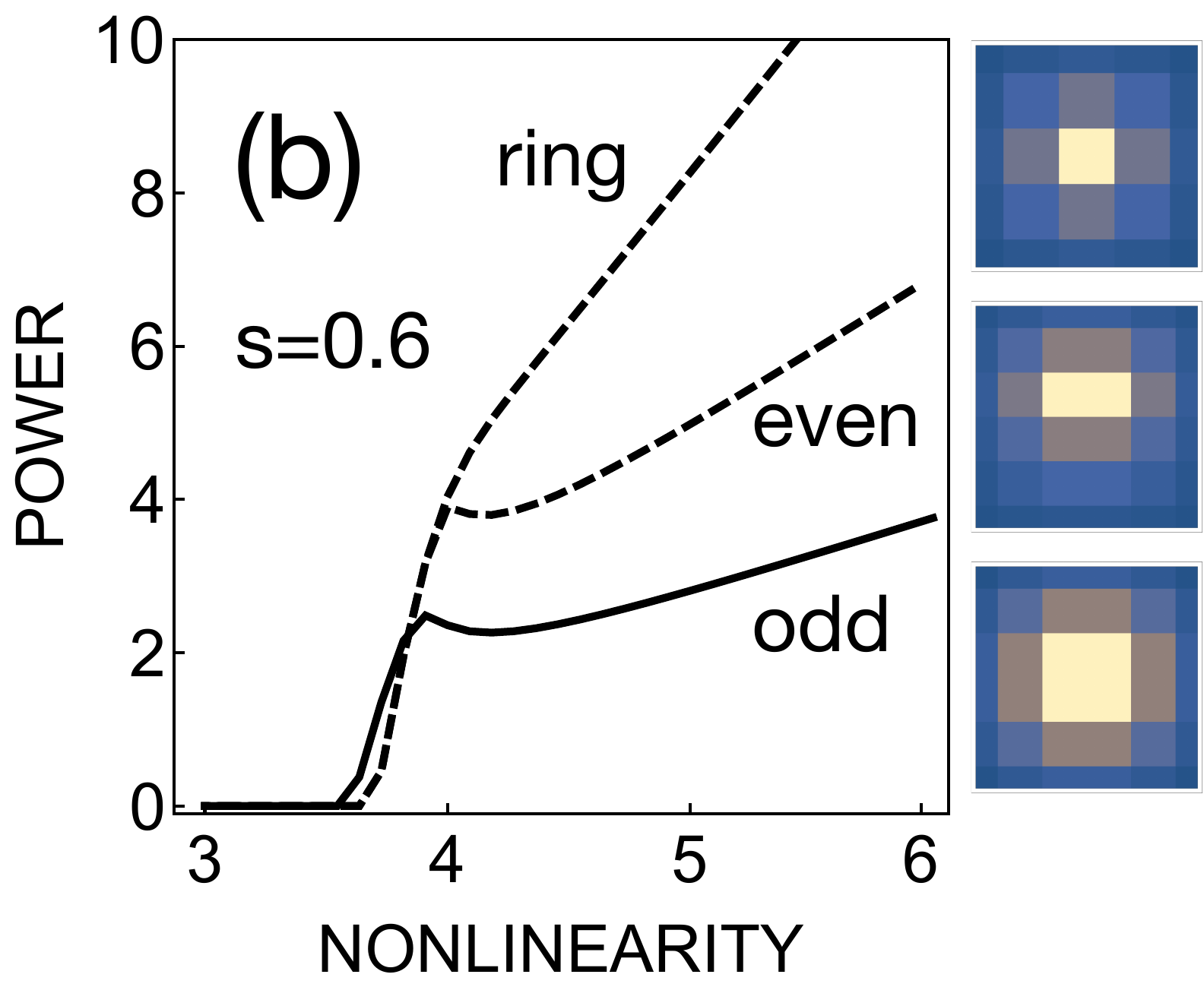}\\
 \includegraphics[scale=0.25]{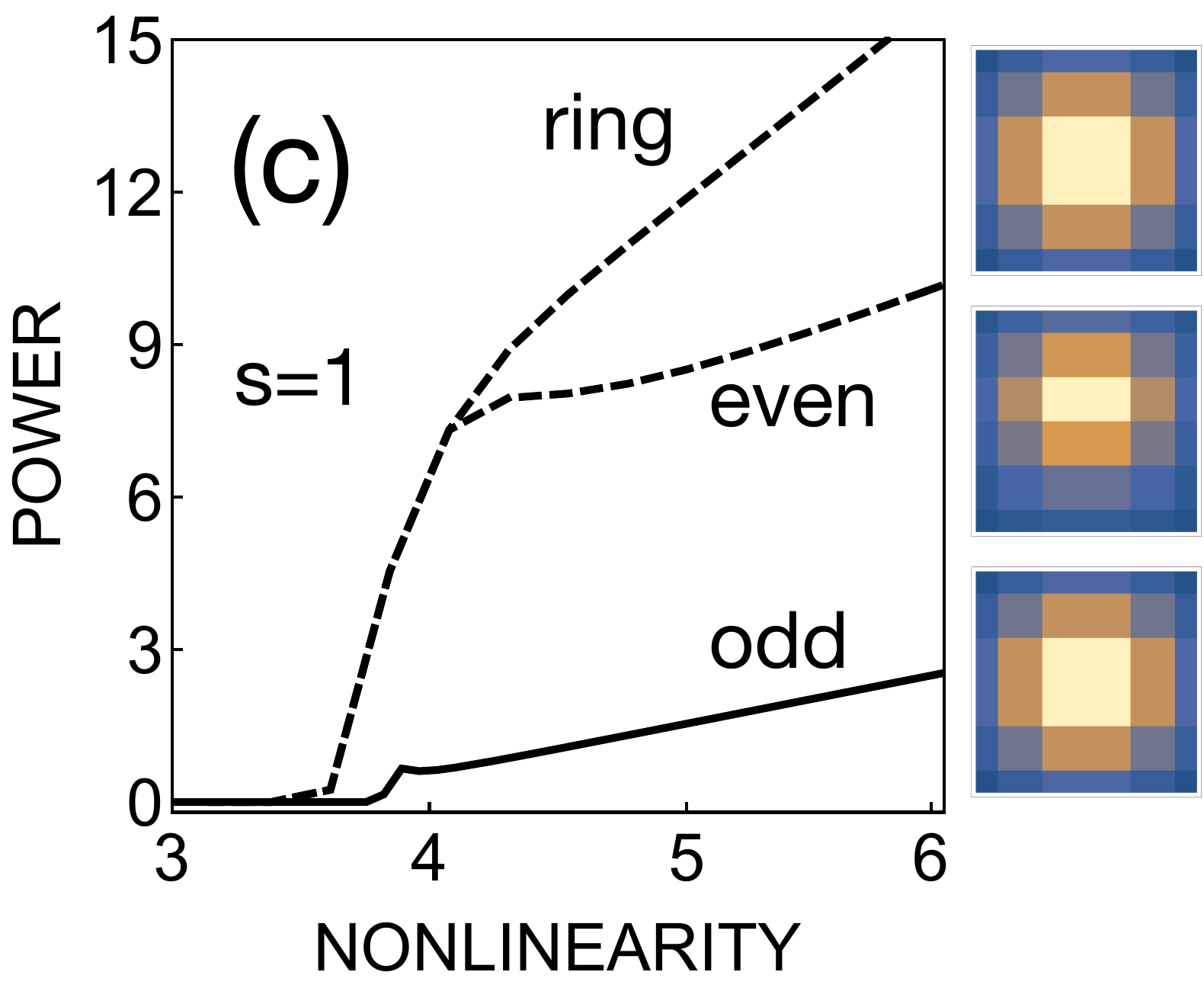}
 \includegraphics[scale=0.25]{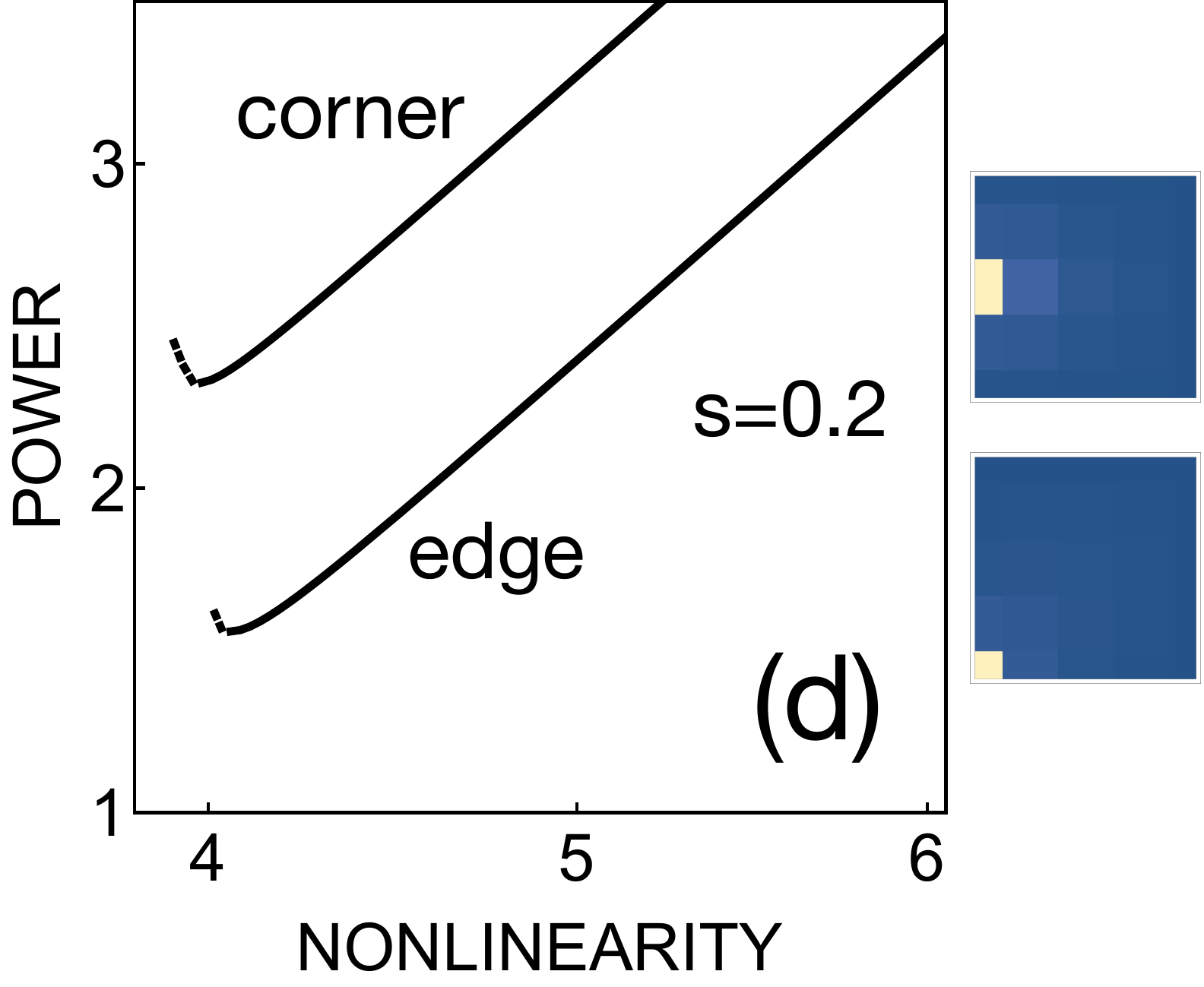}\\
 \includegraphics[scale=0.25]{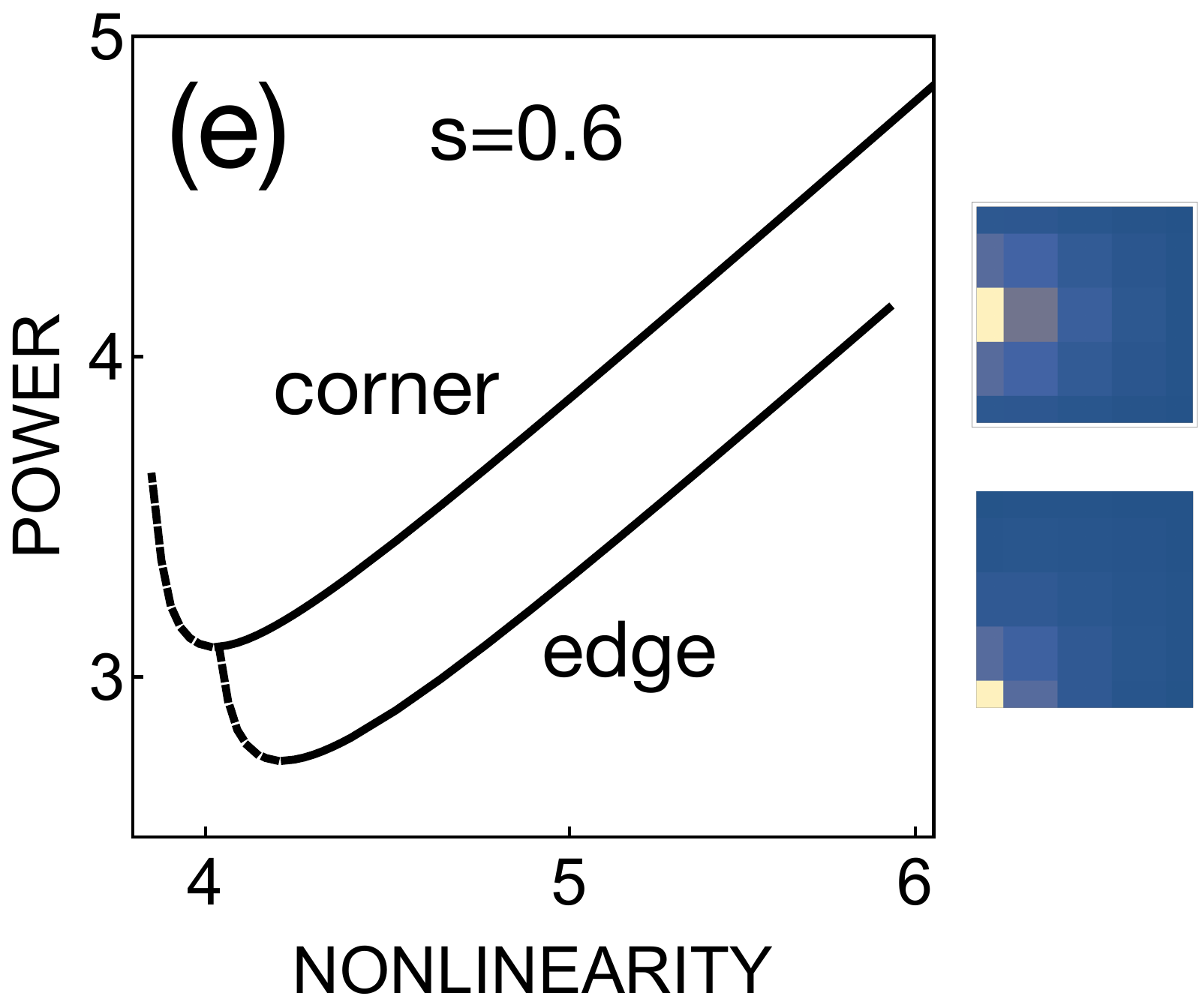}
 \includegraphics[scale=0.25]{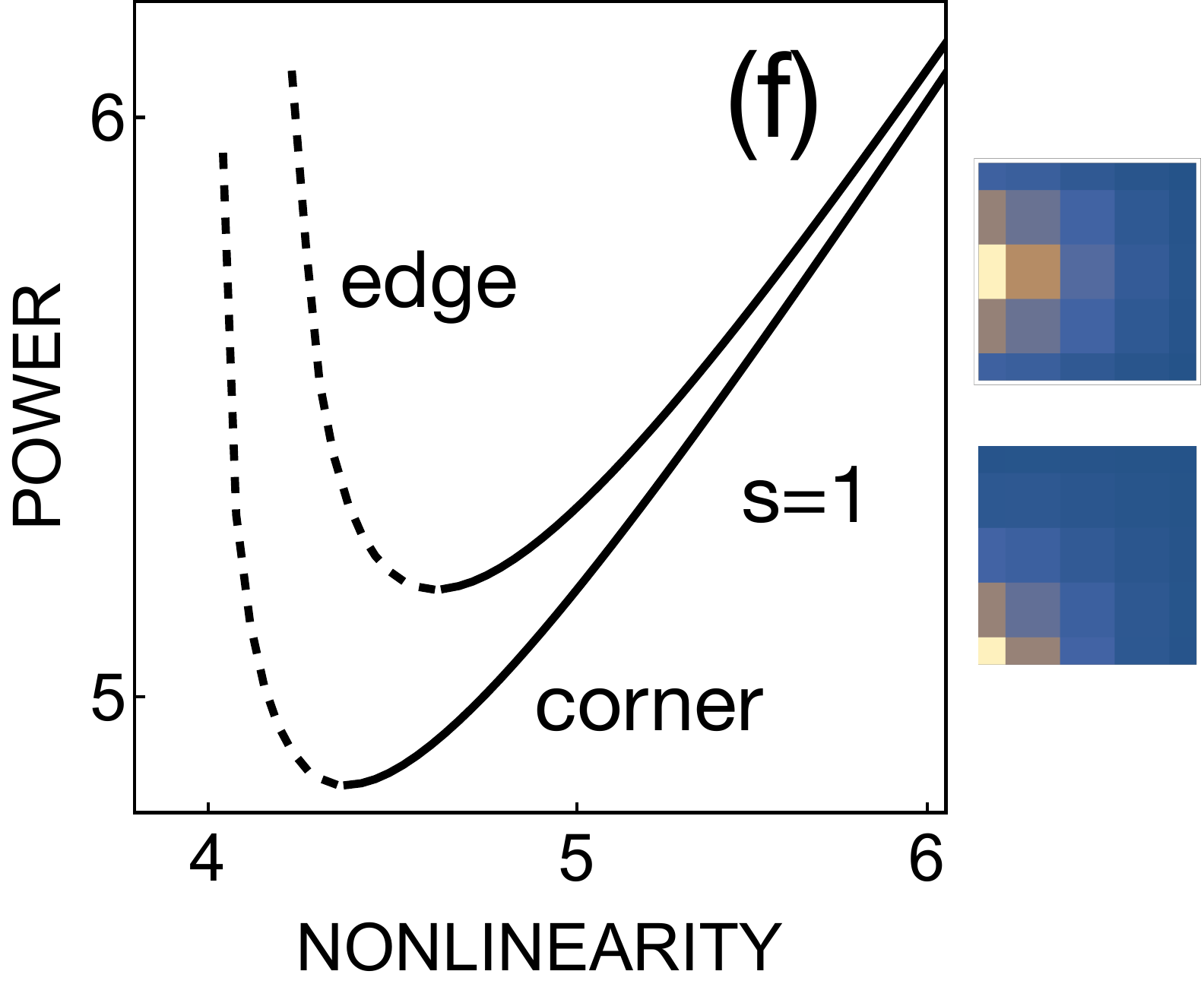}
  \caption{ Power content versus eigenvalue for some bulk and surface modes, for different fractional exponents. Continuous (dashed) curves denote stable (unstable) modes.
   $N=5\times 5$.}
  \label{fig5}
\end{figure}

{\em Nonlinear modes}. Let us consider now nonlinear modes ($\chi\neq 0$), i.e.,  solutions to Eqs.(\ref{stationary}). They constitute a system of $N\times N$ nonlinear algebraic equations, where the form of the nonlinear term adopted here corresponds to Kerr nonlinearity found in coupled waveguide arrays, as well as in the semi classical description of an electron propagating in a deformable lattice. Numerical solutions are obtained by the use of a multidimensional Newton-Raphson scheme, using as a seed the solution obtained from the decoupled limit, also known as the anticontinuous limit. We take a finite $N\times N$ lattice with open boundary conditions and examine two-mode families, ``bulk'' modes, located far from the boundaries and ``surface'' modes located near the beginning (or end) of the lattice. The stability of these nonlinear modes is carried out in the standard manner, which we sketch here for completeness: We perturb our stationary  solution $C_{\bf n}(t) = (\phi_{\bf n}+\delta_{\bf n}(t)) \exp(i \lambda t)$, where $|\delta_{\bf n}(t)|\ll |\phi_{\bf n}|$. We replace this in the evolution equation (\ref{eq:6}) [ with $\Delta_{\bf n}$ replaced by $(\Delta_{\bf n})^s ]$. After a linearization procedure, where we neglect any powers of $\delta_{\bf n}(t)$ beyond the  linear one, we obtain a linear evolution equation for $\delta_{\bf n}(t)$. Next, we decompose $\delta_{\bf n}(t)$ into its real and imaginary parts: $\delta_{\bf n}(t) = x_{\bf n}(t)+ i\ y_{\bf n}(t)$. To simplify the  notation we map the sites of the two-dimensional square lattice into those of an open one-dimensional chain: $(n_{1},n_{2}) \rightarrow n \equiv n_{1} + (n_{2}-1) N$, with $1\leq n_{1}, n_{2}\leq N$. Then, the equations for $x_{n}(t)$ and $y_{n}(t)$ can be written in the form:
\be
{d^2\over{d t^2}}\ \vec{x} - {\bf A}\ {\bf B}\ \vec{x} = 0, \hspace{0.5cm}{d^2\over{d t^2}}\ \vec{y} - {\bf B}\ {\bf A}\ \vec{y} = 0,\label{stability}
\ee
where $\vec{x}=(x_{1}, x_{2},\cdots x_{N\times N})$ and $\vec{y}=(y_{1}, y_{2},\cdots y_{N\times N})$. Matrices ${\bf A}$ and ${\bf B}$ are given by
\be
{\bf A}_{n m} = [-\lambda-K^{s}(0)+\epsilon_{n}+2 |\phi_{n}|^2-\phi_{n}^{2}]\ \delta_{n m}+K^{s}(n-m) \label{20}
\ee
\be
{\bf B}_{n m} =[\lambda+K^{s}(0)-\epsilon_{n}-2 |\phi_{n}|^2-\phi_{n}^2]\ \delta_{n m}-K^{s}(n-m)
\label{21}
\ee
where $\epsilon_{n}=\alpha-\sum_{j}K^{s}(n-j)$, with $\alpha=2$ for a corner site, $\alpha=3$ for an edge site, and $\alpha=4$ for a bulk site. Linear stability is determined by the eigenvalue spectra of the matrices {\bf AB} (or {\bf BA}). When all eigenvalues are real and negative, the system is stable, otherwise it is unstable. In the more general case that considers possible complex eigenvalues, one defines the instability gain $G$ as:
\be
G=\mbox{Max of}\ \left\{ {1\over{2}}\left( \mbox{Re}(g) + \sqrt{\mbox{Re}(g)^2+\mbox{Im}(g)^2}\ \right)\ \right\}^{1/2}
\ee
for all $g$, where $g$ is an eigenvalue of {\bf AB} ({\bf BA}). Thus, when $G=0$, the mode under inspection is stable; otherwise it is unstable. Results from the above procedure are shown in figure \ref{fig5} which shows power versus eigenvalue bifurcation diagrams for some bulk and surface modes, and for several values of the fractional exponents. Also shown are generic shapes of the two-dimensional modes (in this case we have taken a much smaller lattice to reduce computation time) . Since we are dealing with a finite square lattice with open boundaries, there are two surface modes: the `edge' one, and the `corner' one. It is observed (not shown) that, after some few layers below the boundary, the surface modes become almost indistinguishable from the bulk ones. Thus, there is a continuous transition from surface to bulk modes. As for the bulk modes, we have focused on the `odd' mode, centered on a single site, the `even' mode, centered on two nearest-neighbor sites, and the `ring' mode, centered around a closed loop of 4 sites (the two first names originate from the usage employed for one-dimensional lattices).
For the bulk modes, we observe that the fundamental mode is always stable for all $s$ values while the even and ring modes are unstable. All the bulk curves  seem to touch the edge of the linear band, at low powers. As $s$ is decreased, the modes become wider, but there are not other dramatic changes on the shape of the modes. The surface modes decay quickly away from the boundary, but their bifurcations curves look rather similar as $s$ is varied.

\begin{figure}[t]
 \includegraphics[scale=0.22]{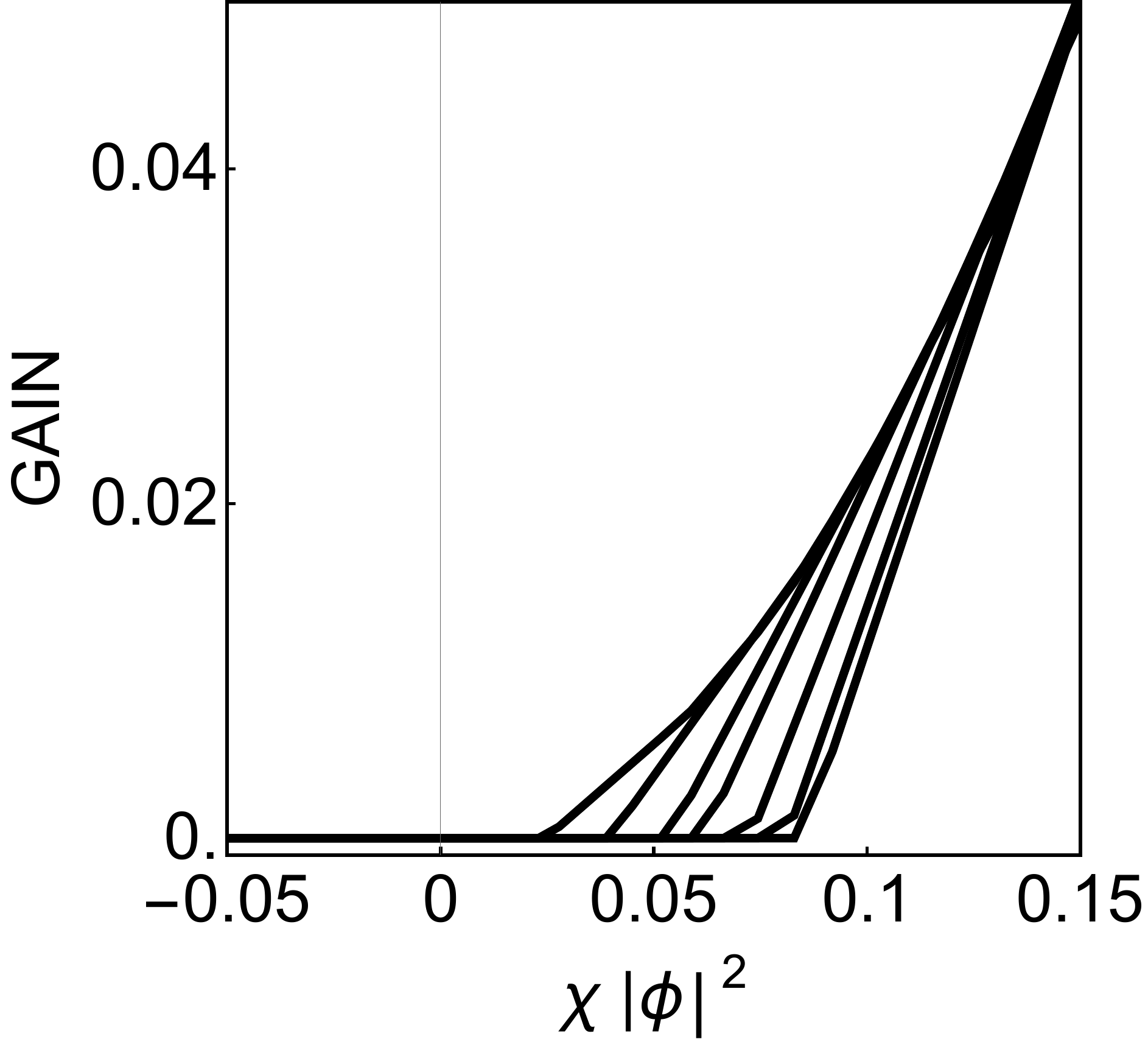}
  \includegraphics[scale=0.21]{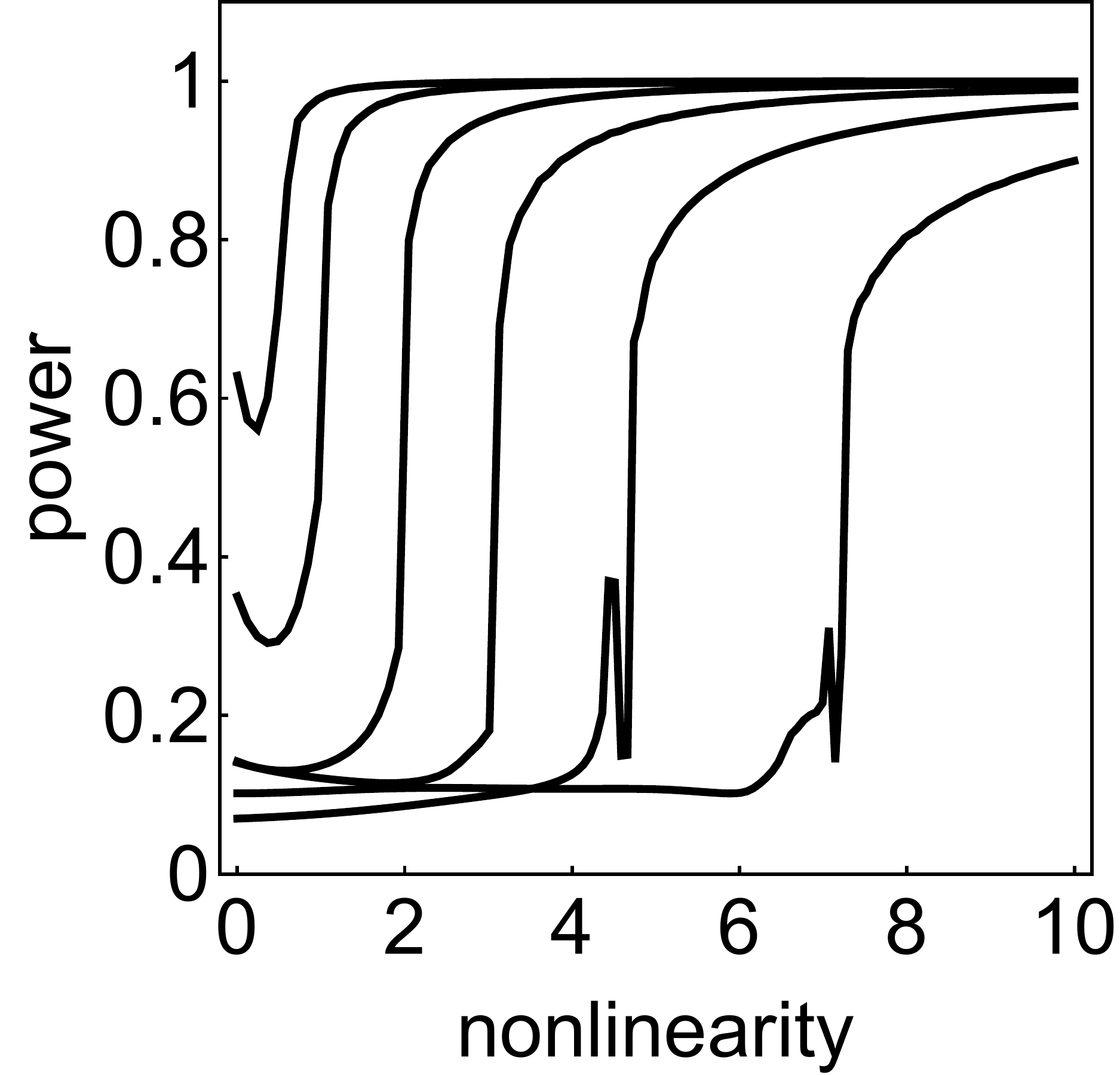}
  \caption{(a) Modulational instability gain versus nonlinearity strength. From the leftmost to the rightmost curve $s = 0.1, 0.2, 0.3, 0.4, 0.6, 0.8, 0.9, 1.0\ (N\times N=25)$ (b) Time-averaged probability at the initial site as a function of nonlinearity, for several different values of the fractional exponent, $s$. From the leftmost to the rightmost curve $s = 0.1, 0.2, 0.4, 0.6, 0.8$ ($T = 20, N\times N=81$).}
  \label{fig6}
\end{figure}

An interesting special case to examine is that of the stability of a nonlinear uniform solution. In one-dimension, the instability of this mode has been observed to give rise to discrete solitons and has, in fact, been proposed as a practical way to produce them\cite{george}. Let us consider a solution of the form $C_{\bf n}=\phi\ \exp(i \lambda t)$. After replacing into the evolution equation one obtains $\lambda = 4 + \chi \phi^2$. Thus, $C_{\bf n}(t)=\phi\ \exp(i(4 + \chi \phi^2))$.
After inserting this form into Eqs.(\ref{20}) and (\ref{21}), one obtains
\be
{\bf A}_{n m} = [ -\sum_{j\neq n}K^{s}(n-j) ]\ \delta_{n m} + K^{s}(n-m) \label{23}
\ee
\be
{\bf B}_{n m} = [ -\sum_{j\neq n}K^{s}(n-j) + 3\ \chi\ \phi^2 ]\ \delta_{n m} + K^{s}(n-m).
\label{24}
\ee
As before, we look at the eigenvalues of ${\bf A}{\bf B}$ (or ${\bf B}{\bf A}$) and record the 
instability gain $G$. Figure \ref{fig6}a shows $G$ as a function of the nonlinearity strength, $\chi |\phi|^2$, for several fractional exponents. We notice that, the stable region ($G=0$) increases with an increase in the fractional exponent $s$. A possible explanation could be that, as $s$ is decreased  form a large value (i.e., near $s=1$), the range of the coupling among sites increases, causing that any perturbation on a given site is instantly felt on distant sites. This situation of mutual, long-range perturbations inhibits the stability of the uniform front, as opposed that case when a perturbation of a given site only affects its immediate neighbors.

{\em Selftrapping}. One the well-known effects of the Kerr nonlinearity is the onset of selftrapping where, for a nonlinearity strength above a threshold value,  an initially localized excitation does not diffuse away completely when placed on a given site of the lattice. After some time, part of the excitation remains localized in the immediate vicinity of the initial site, while the rest diffuse away in a ballistic manner.  We want to examine the possible effect of a fractional exponent on this trapping phenomenon.  To ascertain the presence of a selftrapping transition, one examines the long-time average probability at the initial site (site `zero')
\be
\langle P_{0} \rangle
= {1\over{T}} \int_{0}^{T} |C_{0}(t)|^2\ dt
\ee
We have computed $\langle P_{0} \rangle$ for several $s$ values, comparing the different selftrapping curves. Results are shown in Fig.\ref{fig6}b. We see that, as $s$ is decreased from unity, the position of the selftrapping transition decreases as well and, in the linear limit we notice a degree of linear trapping that increases for lower values of $s$. These results can be explained as follows: as $s$ is decreased, so does the width of the band (Fig.\ref{fig2}). This flattening of the bands originate a smaller group velocity of the modes. On the other hand, the local nonlinearity is roughly equivalent to a linear impurity of strength $\chi |\phi_{0}|^2$. Thus, we have an effective linear impurity embedded in a lattice whose modes have low group velocity (because of relatively flat band). The combination of these two effects, facilitates the trapping of the excitation and thus, decreases the nonlinearity threshold needed. There is yet another effect we can see in Fig.6: As $s$ decreases, the amount of trapping in the limit of zero nonlinearity increases.  This linear trapping approaches unity for $s\rightarrow 0$ and is a consequence of the complete flattening of the band and a complete degeneracy of the modes. This special case has been examined before\cite{simplex1,simplex2}, with the result (adapted to our 2D case) that the time-averaged probability at the initial site is given by
\be
\langle |C_{0}|^2 \rangle = {{(N\times N - 1)^2+1}\over{(N\times N)^2}}
\ee
Therefore, at large $N$ the trapped fraction approaches unity.

{\em Conclusions}.\ We have examined the consequences of employing a fractional version of the usual discrete Laplacian, parametrized by a fractional exponent, on the existence and stability of nonlinear modes, the free propagation of localized linear and nonlinear excitations, and the selftrapping of initially localized excitations, on a two-dimensional square lattice. We found that the main effect of a fractional exponent is to introduce a long-range coupling interaction among the sites of the lattice. The mean 
square displacement is always ballistic with a speed that decreases with a decrease in the fractional exponent. At small values of $s$, one observes a decrease in the bandwidth with a corresponding increase in degeneracy.
The stability of the low-lying excitations is not dissimilar to the one found in the one-dimensional case, while the modulational stability increases with an increase in $s$. Finally, the selftrapping of an initially localized excitation shows a selftrapping transition with a  threshold that increases with $s$, and shows a degree of linear selftrapping at low $s$ values. This was explained to be a consequence of the increase in degeneracy as the band gets flatter and flatter as $s\rightarrow 0$.

The fact that the observed phenomenology observed for this two-dimensional system is not dissimilar to the one found for the one-dimensional analog, points out to the robustness of the discrete soliton phenomenology (nonlinear mode existence and stability, selftrapping, etc) not only against different parameter values of the DNLS equation, but against {\em different mathematical Laplacians}. Also, the fact that the effect of the fractional Laplacian is to introduce a long-range coupling means that one can reproduce experimentally the effect of a fractional Laplacian in an optical context, by setting up an appropriate distribution of refractive indices and inter site distances between waveguides in a waveguide array. Thus, it is possible in principle, to explore the fractional dynamics via optical experiments.

\acknowledgments
The author is grateful to Luz Roncal for valuable discussions.
This work was supported by Fondecyt Grants 1160177 and 1200120.

\end{document}